\documentclass{revtex4}
\usepackage{amssymb}
\usepackage{amsfonts}
\usepackage{amsmath}

\input{tcilatex}

\begin{document}

\title{PDM creation and annihilation operators of the harmonic oscillators
and the emergence of an alternative PDM-Hamiltonian}
\author{Omar Mustafa}
\email{omar.mustafa@emu.edu.tr}
\affiliation{Department of Physics, Eastern Mediterranean University, G. Magusa, north
Cyprus, Mersin 10 - Turkey,\\
Tel.: +90 392 6301378; fax: +90 3692 365 1604.}

\begin{abstract}
\textbf{Abstract:}\ The exact solvability and impressive pedagogical
implementation of the harmonic oscillator's creation and annihilation
operators make it a problem of great physical relevance and the most
fundamental one in quantum mechanics. So would be the position-dependent
mass (PDM) oscillator for the PDM quantum mechanics. We, hereby, construct
the PDM creation and annihilation operators for the PDM oscillator via two
different approaches. First, via von Roos PDM Hamiltonian and show that the
commutation relation between the PDM creation $\hat{A}^{+}$ and annihilation 
$\hat{A}$ operators, $[\hat{A},\hat{A}^{+}]=1\Leftrightarrow $ $\hat{A}\hat{A%
}^{+}-1/2=\hat{A}^{+}\hat{A}+1/2$, not only offers a unique PDM-Hamiltonian $%
\hat{H}_{1}$ but also suggests a PDM-deformation in the coordinate system.
Next, we use a PDM point canonical transformation of the textbook constant
mass harmonic oscillator analog and obtain yet another set of PDM creation $%
\hat{B}^{+}$ and annihilation $\hat{B}$ operators, hence an \emph{%
"apparently new" }\ PDM-Hamiltonian $\hat{H}_{2}$ is obtained. The \emph{%
"new"} PDM-Hamiltonian $\hat{H}_{2}$ turned out to be not only correlated
with $\hat{H}_{1}$ but also represents an alternative and most simplistic
user-friendly PDM-Hamiltonian, $\hat{H}=\left( \hat{p}/\sqrt{2m\left(
x\right) }\right) ^{2}+V\left( x\right) ;$ $\hat{p}=-i\hbar \partial _{x}$,
that has never been reported before.


\end{abstract}

\author{}
\maketitle

\section{Introduction}

The harmonic oscillator problem is one of the most fundamental problems in
classical and quantum mechanics. Its exact solvability and impressive/superb
pedagogical implementation makes it a system of great physical relevance.
Yet, its creation and annihilation operators play an important role in the
build up of its energy basis that are vital components for perturbation
theories treatments. The dynamics of a particle experiencing small
fluctuations near the equilibrium point, $x_{\circ }$ say, allow us to
express the corresponding potential as a Taylor series (i.e. perturbation
series) expansion about $x_{\circ }$. On the other hand, particles with
position-dependent mass (PDM) find their applicability in nuclear physics,
nanophysics, semiconductor, etc \cite{1,2,3,4,5,6,7}. In a more appropriate
and instructive language, it would better be particles with
position-dependent effective mass. That is, a deformation in the coordinate
system may render the mass to be effectively position-dependent. A point
mass moving within the curved coordinates/space transforms into
position-dependent mass in Euclidean coordinates/space (c.f., e.g., \cite%
{8,9,10,11,12} and references cited therein).. Such particles have been
investigated for both classical and/or quantum systems over the years (e.g.,
see the sample of references \cite%
{1,2,3,4,5,6,7,8,9,10,11,12,13,14,15,16,17,18,19,20,21,22,23,24,25,26,27,28,29,30,31,32,33,34,35}%
). It would be interesting, therefore, to put the PDM harmonic oscillator
and its PDM creation and annihilation operators in their pedestal place so
that they find their PDM exact solvability as well as PDM pedagogical
implementation. This would be one of the most fundamental aspects of PDM
quantum mechanics that inspires the content of the current methodical
proposal.

In PDM quantum mechanics, one has to start with the most prominent von Roos
PDM Hamiltonian \cite{1}%
\begin{equation}
\hat{H}=-\frac{1}{4}\left[ M\left( x\right) ^{j}\partial _{x}M\left(
x\right) ^{k}\partial _{x}M\left( x\right) ^{l}+M\left( x\right)
^{l}\partial _{x}M\left( x\right) ^{k}\partial _{x}M\left( x\right) ^{j}%
\right] +V\left( x\right) ,
\end{equation}%
and its ordering ambiguity conflict (an ambiguity that is manifested by the
endless number of its kinetic energy operators that satisfy the von Roos
constraint $j+k+l=-1$). That is, as one changes the values of the ordering
ambiguity parameters $j$, $k$, and $l$ \ through the von Roos constraint,
not only the profile of the kinetic energy operator will change but also the
profile of the effective potential as well. On the theoretically and
physically acceptable sides, nevertheless, it is found that the continuity
conditions at the abrupt heterojunction suggest the parametric condition $%
j=l $ (c.f., e.g., \cite{2,4,7,27}). In the current methodical proposal we
adopt this condition (i.e., $j=l$) and construct the PDM creation and
annihilation operators. Two sets of such operators emerge in the process.
The first of which emerges from the von Roos PDM Hamiltonian (1) and the
other from a PDM point canonical transformation of the textbook constant
mass analog of the harmonic oscillator. Each set of the PDM creation and
annihilation operators results a PDM Hamiltonian that "looks" different from
the other. They turn out to be correlated, nevertheless. We organize our
paper in a sequential order, therefore..

In section 2, we recollect that under the classical PDM settings the force
is no longer given by the time derivative of the linear momentum (i.e., $%
F\neq d\left( M\left( x\right) \dot{x}\right) /dt$) but it is rather given
by $F=\sqrt{M\left( x\right) }d\left( \sqrt{M\left( x\right) }\dot{x}\right)
/dt=-V^{\prime }\left( x\right) $ (c.f., e.g., \cite{18} for more details on
this issue). Consequently, we argue that the interaction potential energy $%
V\left( x\right) $ will be deformed to accommodate the new PDM force
settings. That is, the traditional constant mass harmonic oscillator
potential $V\left( x\right) =m_{\circ }\omega ^{2}x^{2}/2$ is deformed into
a PDM harmonic oscillator potential $V\left( x\right) =m_{\circ }\omega
^{2}Q\left( x\right) \,x^{2}/2$ , where $Q\left( x\right) $ is a PDM
manifested deformation function to be strictly determined in the process of
identifying the PDM creation and annihilation operators. In the first part
of the same section, we use the von Roos PDM Hamiltonian along with a
factorizing recipe to construct PDM creation and annihilation operators, $%
\hat{A}^{+}$ and $\hat{A}$, respectively. Such PDM operators satisfy the
commutation relation $[\hat{A},\hat{A}^{+}]=1$ which consequently implies
that $\hat{A}\hat{A}^{+}-1/2=\hat{A}^{+}\hat{A}+1/2$. The latter is not only
used to build the harmonic oscillator PDM Hamiltonian $\hat{H}_{1}=\omega
\left( \hat{A}\hat{A}^{+}-\frac{1}{2}\right) =\omega \left( \hat{A}^{+}\hat{A%
}+\frac{1}{2}\right) $ as usual, but also we use it to single out one unique
PDM kinetic energy operator. In the second part of the same section, we use
a PDM point canonical transformation of the constant mass textbook analog of
the harmonic oscillator and build up yet an alternative harmonic oscillator
PDM Hamiltonian $\hat{H}_{2}=\omega \left( \hat{B}\,\hat{B}^{+}-\frac{1}{2}%
\right) =\omega \left( \hat{B}^{+}\hat{B}+\frac{1}{2}\right) $ where $\hat{B}%
^{+}$ and $\hat{B}$ are alternative PDM creation and annihilation operators,
satisfying the commutation relation $\left[ \hat{B},\,\hat{B}^{+}\right] =1$%
. The two \emph{"apparently different"} PDM Hamiltonians $\hat{H}_{1}$ and $%
\hat{H}_{2}$ turned out (in the third part of section 2) to be correlated in
such a way that they are alternative forms of each other. A sample of
illustrative examples are given in section 3. Where we use the
PDM-deformation $Q\left( x\right) $ to find the corresponding $m\left(
x\right) $, and the PDM-deformed harmonic oscillator potential (first three
examples). Next, we use the PDM $m\left( x\right) $ to obtain the
corresponding PDM-deformation $Q\left( x\right) $ and the PDM-deformed
harmonic oscillator potential (4th and 5th examples). Finally, we use a
Morse-type PDM-deformed \ harmonic oscillator potential (6th example) and a
Yukawa-type PDM-deformed harmonic oscillator potential (7th example) to find
the corresponding PDM-deformations $Q\left( x\right) $ as well as the
corresponding PDM functions $m\left( x\right) $. \ In our concluding
remarks, section 4, we analyze our results and suggest a new alternative PDM
Hamiltonian (in (70) below) as the most simplistic and user-friendly ever
been reported in the literature. To the best of our knowledge, such results
have never been reported elsewhere.

\section{PDM creation and annihilation harmonic oscillator operators}

In an earlier work. Mustafa \cite{18} has asserted that the force under PDM
settings is no longer given by the time derivative of the linear momentum
(i.e., $F\neq dp\left( x\right) /dt;$ $p\left( x\right) =M\left( x\right) 
\dot{x}$, $M\left( x\right) =m_{\circ }m\left( x\right) $) but it is rather
given by 
\begin{equation}
F=\sqrt{m\left( x\right) }\frac{d}{dt}\left( m_{\circ }\sqrt{m\left(
x\right) }\dot{x}\right) =m_{\circ }m\left( x\right) \ddot{x}+\frac{1}{2}%
m_{\circ }m^{\prime }\left( x\right) \dot{x}^{2}=-V^{\prime }\left( x\right)
;V^{\prime }\left( x\right) =\frac{dV\left( x\right) }{dx},\,\dot{x}=\frac{dx%
}{dt}.
\end{equation}%
For quasi-free PDM particles (i.e., $V\left( x\right) =0$), for example, it
implies that while the PDM pseudo-momentum $\pi \left( x\right) =m_{\circ }%
\sqrt{m\left( x\right) }\dot{x}$ is a conserved quantity, the PDM linear
momentum is no more a conserved quantity. It is, therefore, natural and
convenient to assume that under PDM-settings the harmonic oscillator
potential is deformed in such a way that the constant mass harmonic
oscillator potential $m_{\circ }\omega ^{2}x^{2}/2$ transforms into%
\begin{equation}
V\left( x\right) =\frac{1}{2}m_{\circ }\omega ^{2}Q\left( x\right) \,x^{2}=%
\frac{1}{2}m_{\circ }\omega ^{2}q\left( x\right) ^{2}\text{ };\text{ }%
q\left( x\right) =\sqrt{Q\left( x\right) }x.
\end{equation}%
As long as $q\left( x\right) $ (consequently the PDM manifested deformation $%
Q\left( x\right) $) is to be determined in the process of constructing the
PDM creation $\hat{A}^{+}$ and annihilation $\hat{A}$ operators, this
assumption remains sufficient and\ valid. Moreover, the PDM creation and
annihilation operators should satisfy the commutation relation $[\hat{A},%
\hat{A}^{+}]=1$ as is the case for constant mass settings. Keeping all this
in mind, we construct such PDM operators in two different ways, the first of
which is via the PDM von Roos Hamiltonian (1) and the second is via a PDM
point canonical transformation of their constant mass textbook analog. Two
apparently different PDM Hamiltonians emerge and the correlation between
them is nevertheless identified.

\subsection{via PDM von Roos Hamiltonian}

Enforcing the continuity conditions at the abrupt heterojunction necessarily
implies $j=l$ in (1). Yet, \ the von Roos constraint, $j+k+l=-1$, for $j=l=a$
and $k=2b$ would allow Hamiltonian (1) to collapse into a PDM Hamiltonian
(with $m_{\circ }=\hbar =c=1$ units and $M\left( x\right) =m_{\circ }m\left(
x\right) $ )%
\begin{equation}
\hat{H}_{1}=-\frac{1}{2}m\left( x\right) ^{a}\partial _{x}m\left( x\right)
^{2b}\partial _{x}m\left( x\right) ^{a}+\frac{1}{2}\omega ^{2}\text{ }%
q\left( x\right) ^{2}\text{ ; \ }a+b=-\frac{1}{2},\text{ }q\left( x\right) =%
\sqrt{Q\left( x\right) }x
\end{equation}%
where $V\left( q\left( x\right) \right) =\frac{1}{2}\omega ^{2}q\left(
x\right) ^{2}$ identifies the PDM-deformed oscillator potential and $q\left(
x\right) $ is a position-dependent mass function to be determined in the
sequel (hence, $Q\left( x\right) $ will be determined). However, let us
begin with the construction of the PDM harmonic oscillator creation and
annihilation operators. In so doing, we may appeal to a factorizing recipe
and temporarily suggest that the PDM harmonic oscillator creation operator $%
\hat{A}^{+}$ is given by 
\begin{equation}
\hat{A}^{+}=-\frac{1}{\sqrt{2\omega }}m\left( x\right) ^{a}\partial
_{x}m\left( x\right) ^{b}+\sqrt{\frac{\omega }{2}}\text{ }q\left( x\right) ,
\end{equation}%
and the annihilation operator $\hat{A}$ by 
\begin{equation}
\hat{A}=\frac{1}{\sqrt{2\omega }}m\left( x\right) ^{b}\partial _{x}m\left(
x\right) ^{a}+\sqrt{\frac{\omega }{2}}\text{ }q\left( x\right) ,
\end{equation}%
The harmonic oscillator operators, however, are known to satisfy the
commutation relation%
\begin{equation}
\lbrack \hat{A},\hat{A}^{+}]=1\Longleftrightarrow \hat{A}^{+}\hat{A}+\frac{1%
}{2}=\hat{A}\hat{A}^{+}-\frac{1}{2}.
\end{equation}%
This commutation relation is a necessary and sufficient condition on the
traditional constant mass harmonic oscillator creation and annihilation
operators. So should be the case for the PDM harmonic oscillator creation
and annihilation operators.

Under such settings and in a straightforward manner, one finds that%
\begin{equation}
\hat{A}^{+}\hat{A}=-\frac{1}{2\omega }m\left( x\right) ^{a}\partial
_{x}m\left( x\right) ^{2b}\partial _{x}m\left( x\right) ^{a}-2a\,q\left(
x\right) \left( \frac{1}{\sqrt{m\left( x\right) }}\right) ^{\prime }-\left( 
\frac{q\left( x\right) }{2\sqrt{m\left( x\right) }}\right) ^{\prime }+\frac{%
\omega }{2}q\left( x\right) ^{2}
\end{equation}%
and%
\begin{equation}
\hat{A}\hat{A}^{+}=-\frac{1}{2\omega }m\left( x\right) ^{b}\partial
_{x}m\left( x\right) ^{2a}\partial _{x}m\left( x\right) ^{b}-2a\,q\left(
x\right) \left( \frac{1}{\sqrt{m\left( x\right) }}\right) ^{\prime }-\left( 
\frac{q\left( x\right) }{2\sqrt{m\left( x\right) }}\right) ^{\prime }+\frac{%
\omega }{2}q\left( x\right) ^{2}+\frac{q^{^{\prime }}\left( x\right) }{\sqrt{%
m\left( x\right) }}.
\end{equation}%
The substitution of (8) and (9) in (7) would imply 
\begin{equation}
-\frac{1}{2}m\left( x\right) ^{a}\partial _{x}m\left( x\right) ^{2b}\partial
_{x}m\left( x\right) ^{a}=-\frac{1}{2}m\left( x\right) ^{b}\partial
_{x}m\left( x\right) ^{2a}\partial _{x}m\left( x\right) ^{b}+\frac{q^{\prime
}\left( x\right) }{\sqrt{m\left( x\right) }}-1.
\end{equation}%
This result clearly suggests that the potential related terms vanish to
yield 
\begin{equation}
\frac{q^{\prime }\left( x\right) }{\sqrt{m\left( x\right) }}%
-1=0\Longleftrightarrow q\left( x\right) =\int \sqrt{m\left( x\right) }dx=%
\sqrt{Q\left( x\right) }x,
\end{equation}%
and the kinetic energy terms are equal, i.e.,%
\begin{equation}
-\frac{1}{2}m\left( x\right) ^{a}\partial _{x}m\left( x\right) ^{2b}\partial
_{x}m\left( x\right) ^{a}=-\frac{1}{2}m\left( x\right) ^{b}\partial
_{x}m\left( x\right) ^{2a}\partial _{x}m\left( x\right)
^{b}\Longleftrightarrow a=b.
\end{equation}%
Hence, two basic and critical results are obtained here. The first of which
identifies the form of $q\left( x\right) $ in (11) (hence, relates $Q\left(
x\right) $ with $m\left( x\right) $) and the second restricts the ambiguity
parameters to the identity $a=b$ in (12). Yet, the substitution of $a=b$
into the von Roos constraint $a+b=-1/2$ would result in $a=b=-1/4$.
Consequently, the PDM harmonic oscillator creation (5) and annihilation (6)
operators would, respectively, read 
\begin{equation}
\hat{A}^{+}=-\frac{1}{\sqrt{2\omega }}\frac{1}{\sqrt[4]{m\left( x\right) }}%
\partial _{x}\frac{1}{\sqrt[4]{m\left( x\right) }}+\sqrt{\frac{\omega }{2}}%
q\left( x\right) \Longleftrightarrow \hat{A}^{+}=-i\,\left( \frac{\hat{p}%
\left( x\right) }{\sqrt{2\omega m\left( x\right) }}\right) +\sqrt{\frac{%
\omega }{2}}q\left( x\right) ,
\end{equation}%
and 
\begin{equation}
\hat{A}=\frac{1}{\sqrt{2\omega }}\frac{1}{\sqrt[4]{m\left( x\right) }}%
\partial _{x}\frac{1}{\sqrt[4]{m\left( x\right) }}+\sqrt{\frac{\omega }{2}}%
q\left( x\right) \Longleftrightarrow \hat{A}\left( x\right) =i\,\left( \frac{%
\hat{p}\left( x\right) }{\sqrt{2\omega m\left( x\right) }}\right) +\sqrt{%
\frac{\omega }{2}}q\left( x\right) ,
\end{equation}%
where%
\begin{equation}
\hat{p}\left( x\right) =-i\left( \partial _{x}-\frac{1}{4}\frac{m^{\prime
}\left( x\right) }{m\left( x\right) }\right)
\end{equation}%
is the PDM-momentum operator that has been very recently constructed and
reported by Mustafa and Algadhi \cite{10}. At this point, one may
immediately show that the PDM Hamiltonian $\hat{H}_{1}$ of (4) satisfies the
textbook relation%
\begin{equation}
\hat{H}_{1}=\omega \left( \hat{A}\hat{A}^{+}-\frac{1}{2}\right) =\omega
\left( \hat{A}^{+}\hat{A}+\frac{1}{2}\right) ,
\end{equation}%
and admits its differential form as%
\begin{equation}
\hat{H}_{1}=-\frac{1}{2}\frac{1}{\sqrt[4]{m\left( x\right) }}\partial _{x}%
\frac{1}{\sqrt{m\left( x\right) }}\partial _{x}\frac{1}{\sqrt[4]{m\left(
x\right) }}+\frac{\omega ^{2}}{2}q\left( x\right) ^{2}
\end{equation}%
As such, the so called ambiguity parameters in (4) are no longer ambiguous
but are strictly determined to yield one unique representation for the PDM
kinetic energy operator as%
\begin{equation}
\hat{T}_{1}=-\frac{1}{2}\frac{1}{\sqrt[4]{m\left( x\right) }}\partial _{x}%
\frac{1}{\sqrt{m\left( x\right) }}\partial _{x}\frac{1}{\sqrt[4]{m\left(
x\right) }}=\left( \frac{\hat{p}\left( x\right) }{\sqrt{2m\left( x\right) }}%
\right) ^{2}.
\end{equation}%
Which, in fact, imitates the kinetic energy operator $\hat{T}=\left( \hat{p}/%
\sqrt{2m_{\circ }}\right) ^{2}$ for constant mass settings. At this point,
we may recollect that such parametric ordering in $\hat{T}_{1}$ of (18) is
known in the literature as Mustafa and Mazharimousavi's ordering \cite{27}
(who have used a simple factorization approach for the von Roos PDM
Hamiltonian (1) in general and found that $j=l=-1/4$ and $k=-1/2$). \ Yet,
Cruz et al. \cite{21} have used a supersymmetric approach and geometrically,
shape-wise, compared the corresponding effective potentials (in terms of
superpotentials) with the classical oscillator one ( using different values
for $a$) and found that this PDM kinetic energy operator is graphically and
asymptotically the most suitable ordering.

As a result, we may rewrite the PDM harmonic oscillator Hamiltonian in its
most user-friendly form as%
\begin{equation}
\hat{H}_{1}=\left( \frac{\hat{p}\left( x\right) }{\sqrt{2m\left( x\right) }}%
\right) ^{2}+\frac{\omega ^{2}}{2}q\left( x\right) ^{2}\,;\text{ \ }%
\,q\left( x\right) =\sqrt{Q\left( x\right) }x=\int \sqrt{m\left( x\right) }%
dx.
\end{equation}%
One should notice that the PDM harmonic oscillator creation $\hat{A}^{+}$
and annihilation $\hat{A}$ operators given in terms of the PDM-momentum
operator in (13) and (14) clearly inherit the textbook forms for constant
mass settings, where $m\left( x\right) \longrightarrow m_{\circ }$ and $\hat{%
p}\left( x\right) \longrightarrow \hat{p}=-i\partial _{x}$. Yet, the
commutation relations for constant mass settings are also satisfied by the
PDM settings. That is, one may easily show that%
\begin{equation}
\begin{tabular}{lllll}
$\left[ x,\hat{p}\left( x\right) \right] =i,$ & $\left[ \hat{A},\hat{H}_{1}%
\right] =\hat{A},$ & $\left[ \hat{A}^{+},\hat{H}_{1}\right] =-\hat{A}^{+},$
& $\left[ \hat{A}^{+}\hat{A},\hat{A}\right] =-\hat{A},$ & $\left[ \hat{A}^{+}%
\hat{A},\hat{A}^{+}\right] =\hat{A}^{+},$%
\end{tabular}%
\end{equation}

\subsection{via a PDM point canonical transformation of the constant mass
analog}

Here, we consider a particle with constant mass $m_{\circ }$ moving in the
generalized coordinate $q$ and experiencing a textbook constant mass
harmonic oscillator potential $V\left( q\right) =\frac{1}{2}m_{\circ }\omega
^{2}q^{2}$. The Hamiltonian describing this problem reads (with $m_{\circ
}=\hbar =c=1$ units ) 
\begin{equation}
\hat{H}_{2}=-\frac{1}{2}\partial _{q}^{2}+\frac{1}{2}\omega ^{2}q^{2}
\end{equation}%
We may now recollect that the corresponding textbook creation and
annihilation operators are, respectively, given by

\begin{equation}
\hat{B}^{+}=-\frac{1}{\sqrt{2\omega }}\partial _{q}+\sqrt{\frac{\omega }{2}}%
q,
\end{equation}%
and%
\begin{equation}
\hat{B}=\frac{1}{\sqrt{2\omega }}\partial _{q}+\sqrt{\frac{\omega }{2}}q,
\end{equation}%
where%
\begin{equation}
\hat{H}_{2}=\omega \left( \hat{B}\,\hat{B}^{+}-\frac{1}{2}\right) =\omega
\left( \hat{B}^{+}\hat{B}+\frac{1}{2}\right) \Longleftrightarrow \left[ \hat{%
B},\,\hat{B}^{+}\right] =1.
\end{equation}%
Next, let us use a PDM point canonical transformation in the form of%
\begin{equation}
q=q\left( x\right) =\int \sqrt{m\left( x\right) }dx\Longleftrightarrow dq=%
\sqrt{m\left( x\right) }dx\Longleftrightarrow \partial _{q}=\frac{1}{\sqrt{%
m\left( x\right) }}\partial _{x},
\end{equation}%
similar to (11). This would necessarily transform the constant mass creation
(22) and annihilation (23) operators into PDM creation and annihilation
operators which are, respectively,%
\begin{equation}
\hat{B}^{+}=-\frac{1}{\sqrt{2\omega m\left( x\right) }}\partial _{x}+\sqrt{%
\frac{\omega }{2}}q\left( x\right) ,
\end{equation}%
and%
\begin{equation}
\hat{B}=\frac{1}{\sqrt{2\omega m\left( x\right) }}\partial _{x}+\sqrt{\frac{%
\omega }{2}}q\left( x\right) .
\end{equation}%
Which would, in turn, allow us to write the PDM form of $\hat{H}_{2}$ in
(24) as%
\begin{equation}
\hat{H}_{2}=-\frac{1}{2}\frac{1}{\sqrt{m\left( x\right) }}\partial _{x}\frac{%
1}{\sqrt{m\left( x\right) }}\partial _{x}+\frac{\omega ^{2}}{2}q\left(
x\right) ^{2}\,,\,q\left( x\right) =\sqrt{Q\left( x\right) }x.
\end{equation}

It is obvious and crystal clear that the PDM Hamiltonian $\hat{H}_{1}$ of
(17) (i.e., a von Roos PDM Hamiltonian descendent) and the PDM Hamiltonian $%
\hat{H}_{2}$ of (28) (i.e., a PDM point canonical transformation descendent
of the constant mass textbook analog) are apparently not the same. However,
it is still premature to jump to conclusions at this point, for one has to
check their corresponding eigenvalues and eigenfunctions and perhaps find
out a correlation between them. This is done in the sequel.

\subsection{Correlation between the two PDM Hamiltonians $\hat{H}_{1}$ and $%
\hat{H}_{2}$}

Let us assume that the PDM Hamiltonian $\hat{H}_{1}$ of (17) operates on a
wavefunction $\phi \left( x\right) $ so that the corresponding PDM Schr\"{o}%
dinger equation reads%
\begin{equation}
\hat{H}_{1}\phi \left( x\right) =E_{1}\phi \left( x\right)
\Longleftrightarrow \left\{ -\frac{1}{2}\frac{1}{\sqrt[4]{m\left( x\right) }}%
\partial _{x}\frac{1}{\sqrt{m\left( x\right) }}\partial _{x}\frac{1}{\sqrt[4]%
{m\left( x\right) }}+\frac{\omega ^{2}}{2}q\left( x\right) ^{2}\right\} \phi
\left( x\right) =E_{1}\phi \left( x\right) .
\end{equation}%
On the other hand, the PDM Hamiltonian $\hat{H}_{2}$ of (28) is assumed to
operate on a wavefunction $\Psi \left( q\right) =\Psi \left( q\left(
x\right) \right) $ so that the corresponding PDM Schr\"{o}dinger equation
reads%
\begin{equation}
\hat{H}_{2}\Psi \left( q\left( x\right) \right) =E_{2}\Psi \left( q\left(
x\right) \right) \Longleftrightarrow \left\{ -\frac{1}{2}\frac{1}{\sqrt{%
m\left( x\right) }}\partial _{x}\frac{1}{\sqrt{m\left( x\right) }}\partial
_{x}+\frac{\omega ^{2}}{2}q\left( x\right) ^{2}\right\} \Psi \left( q\left(
x\right) \right) =E_{2}\Psi \left( q\left( x\right) \right) .
\end{equation}%
Our objective here is to bring the PDM Schr\"{o}dinger equation (29) into a
similar form as that of (30). In so doing, let us multiply (29), from the
left, by $1/\sqrt[4]{m\left( x\right) }$ and rearrange terms to get%
\begin{equation}
\left\{ -\frac{1}{2}\frac{1}{\sqrt{m\left( x\right) }}\partial _{x}\frac{1}{%
\sqrt{m\left( x\right) }}\partial _{x}+\frac{\omega ^{2}}{2}q\left( x\right)
^{2}\right\} m\left( x\right) ^{-1/4}\phi \left( x\right) =E_{1}m\left(
x\right) ^{-1/4}\phi \left( x\right) .
\end{equation}%
Therefore, if one demands that the PDM Hamiltonian $\hat{H}_{1}$ of (17) is
isospectral with the PDM Hamiltonian $\hat{H}_{2}$ of (28) (which is indeed
the case here, for we have the very same PDM harmonic oscillator problem for
both Hamiltonians) we may then conclude that%
\begin{equation}
\Psi \left( q\left( x\right) \right) =m\left( x\right) ^{-1/4}\phi \left(
x\right) \Longleftrightarrow E_{1}=E_{2}=E.
\end{equation}%
Consequently, we may now safely argue that%
\begin{equation}
\hat{H}_{2}\Psi \left( q\left( x\right) \right) =m\left( x\right) ^{-1/4}%
\hat{H}_{1}\phi \left( x\right) =E\Psi \left( q\left( x\right) \right)
=Em\left( x\right) ^{-1/4}\phi \left( x\right) .
\end{equation}%
Then, the correlation between the two sets of PDM creation and annihilation
operators now reads%
\begin{equation}
\left( \hat{A}\,\hat{A}^{+}-\frac{1}{2}\right) \phi \left( x\right) =m\left(
x\right) ^{1/4}\left( \hat{B}\,\hat{B}^{+}-\frac{1}{2}\right) \Psi \left(
q\left( x\right) \right) .
\end{equation}%
The connection and mapping between the two PDM systems is made clear,
therefore.

At this point, however, the reader should be aware of the fact that the
eigenvalues $E$ and eigenfunctions $\Psi \left( q\left( x\right) \right) $
are nothings but the exact textbook harmonic oscillator's and are ,
respectively, given by%
\begin{equation}
E_{n}=\omega \left( n+\frac{1}{2}\right) \text{ \ , \ }\Psi _{n}\left(
q\left( x\right) \right) =\mathcal{N}_{n}\,\exp \left( -\frac{q\left(
x\right) ^{2}}{2}\right) \,H_{n}\left( q\left( x\right) \right) \text{ };%
\text{ }n=0,1,2,\cdots ,
\end{equation}%
where%
\begin{equation}
H_{n}\left( q\left( x\right) \right) =H_{n}\left( q\right) =\left( -1\right)
^{n}\exp \left( q^{2}\right) \,\frac{d^{n}}{dq^{n}}\exp \left( -q^{2}\right)
\end{equation}%
are the Hermit polynomials and $n$ represents the principle quantum number.
Yet, both $\hat{H}_{2}\Psi \left( q\left( x\right) \right) $ and $\hat{H}%
_{1}\phi \left( x\right) $ would result a textbook like PDM Schr\"{o}dinger
equation in the form%
\begin{equation}
\left\{ \left( \frac{\hat{p}\left( x\right) }{\sqrt{2m\left( x\right) }}%
\right) ^{2}+\frac{\omega ^{2}}{2}q\left( x\right) ^{2}\right\} \,\phi
_{n}\left( x\right) =E_{n}\,\phi _{n}\left( x\right) ,
\end{equation}%
which, indeed, not only replicates the traditional Schr\"{o}dinger equation
but also collapses exactly into that for constant mass settings. Moreover,
the eigenvalues and eigenfunctions are readily exactly known for all
integrable $\sqrt{m\left( x\right) }$ of $q\left( x\right) $ in (11). A
sample of illustrative examples is given below.

\section{A sample of illustrative examples}

Apriori, wee have asserted that $q\left( x\right) $ of (11) determines the
relation between $Q\left( x\right) $ and $m\left( x\right) $. That is, while
the second part of (11) yields that 
\begin{equation}
q\left( x\right) =\sqrt{Q\left( x\right) }x\Longleftrightarrow q^{\prime
}\left( x\right) =\frac{dq\left( x\right) }{dx}=\sqrt{Q\left( x\right) }%
\left( 1+\frac{Q^{\prime }\left( x\right) }{2\,Q\left( x\right) }x\right) ,
\end{equation}%
the first part of (11), on the other hand, implies (compared with (38)) that%
\begin{equation}
q^{\prime }\left( x\right) =\sqrt{m\left( x\right) }\Longleftrightarrow 
\sqrt{m\left( x\right) }=\sqrt{Q\left( x\right) }\left( 1+\frac{Q^{\prime
}\left( x\right) }{2\,Q\left( x\right) }x\right) .
\end{equation}%
This result would determine the form of $q\left( x\right) $ in (35) and
consequently the form of the PDM harmonic oscillator potential%
\begin{equation}
V\left( x\right) =\frac{\omega ^{2}}{2}Q\left( x\right) \,x^{2}.
\end{equation}%
where $Q\left( x\right) $ is a dimensionless scalar multiplier that
represents, hereinafter, a PDM-deformation function (introduced as a
manifestation of the PDM setting) of the constant mass harmonic oscillator
potential $\,\omega ^{2}x^{2}/2$. Therefore, our potential $V\left( x\right) 
$ in (40)\ represents a PDM-deformed harmonic oscillator potential.\ Yet at
this point, it should be noted that $Q\left( x\right) =const.$ and $Q\left(
x\right) =m\left( x\right) $ would immediately retrieve the constant mass
setting, and shall not be discussed here, therefore. Under such PDM
implications, the PDM harmonic oscillator potential can never be expressed
as $V\left( x\right) =m\left( x\right) \omega ^{2}x^{2}/2$ but rather it
should be expressed as in (40) where condition (39) determines the nature of
the relation between PDM-deformation function $Q\left( x\right) $ and $%
m\left( x\right) $ (c.f., e.g., Cruz cruz et al. \cite{21} and Carinena et
al. \cite{34}).

Apart from the constant mass setting, we now consider the following sample
of illustrative examples. All of which admit exact eigenvalues and
eigenfunctions inherited from our results in (33), (35), and (36) as the
exact solutions for the PDM Schr\"{o}dinger equation in (37). They are
given, respectively, as 
\begin{equation}
E_{n}=\omega \left( n+\frac{1}{2}\right) \text{, \ }\phi _{n}\left( x\right)
=m\left( x\right) ^{1/4}\Psi _{n}\left( q\left( x\right) \right) ,\text{\ }%
\Psi _{n}\left( q\left( x\right) \right) =\mathcal{N}_{n}\,\exp \left( -%
\frac{q\left( x\right) ^{2}}{2}\right) \,H_{n}\left( q\left( x\right)
\right) ;\,\,n=0,1,2,\cdots ,
\end{equation}%
provided that $Q\left( x\right) $ (or equivalently $m\left( x\right) $), is
determined through (39). This is not only restricted to the sample of
illustrative examples below but also for every $Q\left( x\right) $ and $%
m\left( x\right) $ satisfying (39) provided that they are, mathematical
and/or quantum mechanical wise, well-behaved functions (see, for example,
the sample of PDM-functions in \cite{21,34}, some of which are used below) .

\subsection{A PDM sample case without singularities $m\left( x\right)
=\left( 1+\protect\lambda x^{2}\right) ^{-3}$}

The assumption that the PDM deformation function $Q\left( x\right) $ is
given by%
\begin{equation}
Q\left( x\right) =\frac{1}{1+\lambda x^{2}}
\end{equation}%
without singularities, would allow us to obtain, through (39), a PDM without
singularities as%
\begin{equation}
m\left( x\right) =\frac{1}{\left( 1+\lambda x^{2}\right) ^{3}}%
\Longleftrightarrow q\left( x\right) =\frac{x}{\sqrt{1+\lambda x^{2}}}.
\end{equation}%
Consequently, the PDM-deformed harmonic oscillator potential (40) reads%
\begin{equation}
V\left( x\right) =\frac{\omega ^{2}}{2}\left( \frac{x^{2}}{1+\lambda x^{2}}%
\right) ,
\end{equation}%
to admit the exact eigenvalues and eigenfunctions in (41) as the solutions
of the PDM-Schr\"{o}dinger equation (37) with $q\left( x\right) $ of (43).

\subsection{A PDM sample case with singularities $m\left( x\right) =\left( 
\protect\lambda x^{2}-1\right) ^{-3}$}

If we consider that the PDM-deformation function $Q\left( x\right) $ has
singularities and given by 
\begin{equation}
Q\left( x\right) =\frac{1}{\lambda x^{2}-1},
\end{equation}%
then a PDM with singularities is obtained as 
\begin{equation}
m\left( x\right) =\frac{1}{\left( \lambda x^{2}-1\right) ^{3}}%
\Longleftrightarrow q\left( x\right) =\frac{x}{\sqrt{\lambda x^{2}-1}},
\end{equation}%
and the corresponding PDM-deformed harmonic oscillator potential (40) reads%
\begin{equation}
V\left( x\right) =\frac{\omega ^{2}}{2}\left( \frac{x^{2}}{\lambda x^{2}-1}%
\right) \,.
\end{equation}%
with the exact eigenvalues and eigenfunctions in (41) as the solutions of
the PDM-Schr\"{o}dinger equation (37) with $q\left( x\right) $ of (43).

\subsection{A power-law PDM sample case $m\left( x\right) \sim x^{\protect%
\sigma }$}

A power-law PDM-deformation function%
\begin{equation}
Q\left( x\right) =\lambda x^{\sigma },
\end{equation}%
would lead to a PDM function%
\begin{equation}
m\left( x\right) =\left( 1+\frac{\sigma }{2}\right) ^{2}\lambda x^{\sigma
}\Longleftrightarrow q\left( x\right) =\sqrt{\lambda }x^{\left( \sigma
+2\right) /2},
\end{equation}%
where $\sigma \neq -2,0;\sigma \in 
\mathbb{N}
$, otherwise trivial solutions or constant mass setting are, respectively,
manifested. Under such settings, a power-law type PDM-deformed harmonic
oscillator potential (40) is obtained as%
\begin{equation}
V\left( x\right) =\frac{\omega ^{2}}{2}\lambda x^{\sigma +2};\text{ }\sigma
\neq -2,0;\sigma \in 
\mathbb{N}
.
\end{equation}%
That admits the exact eigenvalues and eigenfunctions of (41) as the
solutions of the PDM-Schr\"{o}dinger equation (37) with $q\left( x\right) $
of (49).

\subsection{A PDM sample case without singularities $m\left( x\right)
=\left( 1+\protect\alpha ^{2}x^{2}\right) ^{-1}$}

We now start with a PDM function $m\left( x\right) $ that has no
singularities \cite{21,34} as%
\begin{equation}
m\left( x\right) =\frac{1}{\alpha ^{2}x^{2}+1}.
\end{equation}%
in (11) would imply that%
\begin{equation}
Q\left( x\right) =\left[ \frac{\ln \left( \alpha x+\sqrt{\alpha ^{2}x^{2}+1}%
\right) +\beta }{\alpha x}\right] ^{2}\Longleftrightarrow q\left( x\right) =%
\frac{1}{\alpha }\left[ \ln \left( \alpha x+\sqrt{\alpha ^{2}x^{2}+1}\right)
+\beta \right] .
\end{equation}%
Consequently a logarithmic-type PDM-deformed harmonic oscillator potential
(40) is obtained as%
\begin{equation}
V\left( x\right) =\frac{\omega ^{2}}{2\alpha ^{2}}\left[ \ln \left( \alpha x+%
\sqrt{\alpha ^{2}x^{2}+1}\right) +\beta \right] ^{2},
\end{equation}%
and admits the exact eigenvalues and eigenfunctions of (41) as the solutions
of the PDM-Schr\"{o}dinger equation (37) with $q\left( x\right) $ of (52).

\subsection{A PDM sample case with singularities $m\left( x\right) =\left( 1-%
\protect\alpha ^{2}x^{2}\right) ^{-2}$}

A PDM function $m\left( x\right) $ with singularities \cite{21,34}%
\begin{equation}
m\left( x\right) =\frac{1}{\left( 1-\alpha ^{2}x^{2}\right) ^{2}}
\end{equation}%
in (11), yields%
\begin{equation}
Q\left( x\right) =\frac{1}{4\alpha ^{2}x^{2}}\left[ \ln \left( \frac{\alpha
x-1}{\alpha x+1}\right) +\beta \right] ^{2}\Longleftrightarrow q\left(
x\right) =\frac{1}{2\alpha }\left[ \ln \left( \frac{\alpha x-1}{\alpha x+1}%
\right) +\beta \right]
\end{equation}%
and a logarithmic-type PDM-deformed harmonic oscillator potential (40) is
obtained as%
\begin{equation}
V\left( x\right) =\frac{\omega ^{2}}{8\alpha ^{2}}\left[ \ln \left( \frac{%
\alpha x-1}{\alpha x+1}\right) +\beta \right] ^{2},
\end{equation}%
with its exact eigenvalues and eigenfunctions in (41) as the solutions of
the PDM-Schr\"{o}dinger equation (37) with $q\left( x\right) $ of (55).

\subsection{A PDM in a Morse-type potential $V\left( x\right) =A\left( e^{-2%
\protect\beta x}-2e^{-\protect\beta x}\right) $}

A Morse-type PDM-deformed harmonic oscillator potential (40)%
\begin{equation}
V\left( x\right) =\frac{\lambda \omega ^{2}}{2}\left( e^{-2\beta
x}-2e^{-\beta x}\right) ,
\end{equation}%
would suggest that a PDM-deformation $Q\left( x\right) $ is given in the
form of%
\begin{equation}
Q\left( x\right) =\frac{\lambda }{x^{2}}\left( e^{-2\beta x}-2e^{-\beta
x}\right) \Longleftrightarrow q\left( x\right) =\sqrt{\lambda \left(
e^{-2\beta x}-2e^{-\beta x}\right) },
\end{equation}%
with a PDM function%
\begin{equation}
m\left( x\right) =\lambda \beta ^{2}\frac{\left( e^{-\beta x}-1\right) ^{2}}{%
1-2e^{-\beta x}}.
\end{equation}%
The corresponding exact eigenvalues and eigenfunctions in (41) as the
solutions of the PDM-Schr\"{o}dinger equation (37) with $q\left( x\right) $
of (57).

\subsection{A PDM in Yukawa-type potential $V\left( x\right) =-Be^{-\protect%
\delta x}/x$}

A Yukawa-type PDM-deformed harmonic oscillator potential (40) 
\begin{equation}
V\left( x\right) =-\frac{V_{\circ }\omega ^{2}}{2}\left( \frac{e^{-\delta x}%
}{x}\right)
\end{equation}%
would imply a PDM-deformation function%
\begin{equation}
Q\left( x\right) =-V_{\circ }\frac{e^{-\delta x}}{x^{3}}\Longleftrightarrow
q\left( x\right) =\sqrt{-V_{\circ }\frac{e^{-\delta x}}{x}},
\end{equation}%
and a PDM function%
\begin{equation}
m\left( x\right) =-\frac{V_{\circ }}{4}\left( \frac{\delta x+1}{x^{3}}%
\right) e^{-\delta x}.
\end{equation}%
The exact eigenvalues and eigenfunctions of which are given in (41) as the
solutions of the PDM-Schr\"{o}dinger equation (37) with $q\left( x\right) $
of (60).

\section{Concluding Remarks}

In this work, we have constructed the PDM creation and annihilation
operators for the PDM-deformed harmonic oscillators through two different
ways: (i) via von Roos PDM-Hamiltonian (4), and (ii) via a PDM point
canonical transformation (25) of the textbook constant mass Hamiltonian
analog (21). Using the von Roos Hamiltonian (4), we have shown that the
commutation relation (7) (between the PDM creation $\hat{A}^{+}$ and
annihilation $\hat{A}$ operators for the PDM-deformed harmonic oscillators)
offers a strict determination of the PDM kinetic energy operator (18) and
suggests a PDM-deformation in the coordinate system (11) (consequently, a
PDM-deformation $Q\left( x\right) $ in the PDM-deformed harmonic oscillator
potential (3)) to imply the PDM Hamiltonian $\hat{H}_{1}$ of (17). On the
other hand, the PDM point canonical transformation (25) yields another set
of PDM creation $\,\hat{B}^{+}$ and annihilation $\hat{B}\,$operators, (26)
and (27), for the PDM-deformed harmonic oscillators to result yet another
PDM Hamiltonian $\hat{H}_{2}$ of (28). In the process, therefore, two
"apparently" different PDM Hamiltonian operators (or equivalently, two
different PDM kinetic energy operators) are obtained. In the literature,
however, $\hat{H}_{1}$ is known to represent Mustafa and Mazharimousavi's
parametric ordering \cite{27} (i.e., $j=\ell =-1/4$ and $k=-1/2$, obtained
through the factorization of the kinetic energy term of (1)). Whereas, $\hat{%
H}_{2}$ turned out to be correlated with $\hat{H}_{1}$, through (33), but
has never been reported as a feasible ordering in the literature before. As
a result, one PDM Hamiltonian, $\hat{H}_{1}$, is effectively and vividly
singled out of the von Roos PDM Hamiltonian (1). The corresponding PDM-Schr%
\"{o}dinger equation of which, in general, reads%
\begin{equation}
\left\{ \left( \frac{\hat{p}\left( x\right) }{\sqrt{2m\left( x\right) }}%
\right) ^{2}+V\left( q\left( x\right) \right) \right\} \,\phi _{n}\left(
x\right) =E_{n}\,\phi _{n}\left( x\right) ,\text{ \ }
\end{equation}%
for any PDM-deformed interaction potential field $V(q\left( x\right) )$.
Where, $\hat{p}\left( x\right) =-i\left( \partial _{x}-\frac{1}{4}m^{\prime
}\left( x\right) /m\left( x\right) \right) .$is the PDM-momentum operator
very recently reported by Mustafa and Algadhi \cite{10}, and%
\begin{equation}
q\left( x\right) =\sqrt{Q\left( x\right) }x\text{ },\text{ }\sqrt{m\left(
x\right) }=\sqrt{Q\left( x\right) }\left( 1+\frac{Q^{\prime }\left( x\right) 
}{2\,Q\left( x\right) }x\right) .
\end{equation}%
For the PDM-deformed harmonic oscillator $V\left( q\left( x\right) \right) =%
\frac{1}{2}\omega ^{2}q\left( x\right) ^{2}=\frac{1}{2}\omega ^{2}Q\left(
x\right) x^{2}$, for example, the PDM creation and annihilation operators
are constructed as in (13) and (14), respectively.

Obviously, the PDM-Schr\"{o}dinger equation (63) as well as the PDM creation
and annihilation operators, (13) and (14), look very much like their
textbook counterparts for constant mass settings. Yet, they satisfy the
textbook commutation relations (20). The correlation%
\begin{equation*}
\left( \hat{A}\,\hat{A}^{+}-\frac{1}{2}\right) \phi \left( x\right) =m\left(
x\right) ^{1/4}\left( \hat{B}\,\hat{B}^{+}-\frac{1}{2}\right) \Psi \left(
q\left( x\right) \right) ,
\end{equation*}%
on the other hand, allows one to build up the energy eigenvalues and
eigenfunctions for PDM-deformed harmonic oscillator potential. That is, one
may (in the Dirac notations) use 
\begin{equation}
\hat{B}\,\left/ \,\Psi _{n}\left( q\left( x\right) \right) \right\rangle =%
\sqrt{n}\,\left/ \,\Psi _{n-1}\left( q\left( x\right) \right) \right\rangle 
\text{ ; \ }\hat{B}^{+}\left/ \,\Psi _{n}\left( q\left( x\right) \right)
\right\rangle =\sqrt{n+1}\,\left/ \,\Psi _{n+1}\left( q\left( x\right)
\right) \right\rangle \,;\,n=0,1,2,\cdots ,
\end{equation}%
to build up the spectrum through 
\begin{equation}
E_{n}=\omega \left( n+\frac{1}{2}\right) ,
\end{equation}%
and%
\begin{equation}
\Psi _{n}\left( q\left( x\right) \right) =m\left( x\right) ^{-1/4}\phi
_{n}\left( x\right) =\mathcal{N}_{n}\exp \left( -\frac{q\left( x\right) ^{2}%
}{2}\right) \,H_{n}\left( q\left( x\right) \right) .
\end{equation}%
In our illustrative examples (section 3), moreover, we have used the
PDM-deformation $Q\left( x\right) $ to find the corresponding $q\left(
x\right) $, $m\left( x\right) $, and the PDM-deformed harmonic oscillator
potential (in the first three examples). Next, we have used the PDM $m\left(
x\right) $ to obtain the corresponding PDM-deformation $Q\left( x\right) $, $%
q\left( x\right) $, and the PDM-deformed harmonic oscillator potential (in
the 4th and 5th examples). Then, we have used a Morse-type PDM-deformed \
harmonic oscillator potential (6th example) and a Yukawa-type PDM-deformed
harmonic oscillator potential (7th example) to find the corresponding
PDM-deformations $Q\left( x\right) $ as well as the corresponding PDM
functions $m\left( x\right) $.

In the light of our experience through the current methodical proposal, a
critical and new observation is unavoidable. The fact that the PDM
Hamiltonians $\hat{H}_{1}$ and $\hat{H}_{2}$ are isospectral, (32), and are
correlated through (33) immediately suggests that%
\begin{equation}
\hat{H}_{2}=-\frac{1}{2}\frac{1}{\sqrt{m\left( x\right) }}\partial _{x}\frac{%
1}{\sqrt{m\left( x\right) }}\partial _{x}+V\left( q\left( x\right) \right)
\,,\,q\left( x\right) =\sqrt{Q\left( x\right) }x,
\end{equation}%
is yet an alternative PDM Hamiltonian which does not belong to the set of
von Roos PDM Hamiltonians of (1). Moreover, as long as the eigenvalues and
eigenfunctions are to be determined, then $\hat{H}_{2}$ of (68) should,
hereinafter, be realized to be yet another equivalent/alternative and viable
PDM Hamiltonian. Which, for the PDM-deformed harmonic oscillator discussed
above, offers a straightforward construction of the eigenvalues and
eigenfunction as well as a new set of PDM creation $\,\hat{B}^{+}$ and
annihilation $\hat{B}\,$operators, (26) and (27), that can be expressed in
terms of the regular constant mass momentum operator $\hat{p}=-i\partial
_{x} $, respectively, as%
\begin{equation}
\hat{B}^{+}=-i\frac{\hat{p}}{\sqrt{2\omega m\left( x\right) }}+\sqrt{\frac{%
\omega }{2}}q\left( x\right) \text{ \ , \ \ \ }\hat{B}=i\frac{\hat{p}}{\sqrt{%
2\omega m\left( x\right) }}+\sqrt{\frac{\omega }{2}}q\left( x\right) \text{
\ ; \ }\hat{p}=-i\partial _{x}.
\end{equation}%
Under such new PDM settings, one may now write, in general, the PDM-Schr\"{o}%
dinger equation corresponding to $\hat{H}_{2}$ of (68) as%
\begin{equation}
\left\{ \left( \frac{\hat{p}}{\sqrt{2m\left( x\right) }}\right) ^{2}+V\left(
q\left( x\right) \right) \right\} \Psi _{n}\left( q\left( x\right) \right)
=E_{n}\,\Psi _{n}\left( q\left( x\right) \right) ,
\end{equation}%
where, $q\left( x\right) $ is given by (64). Finally, all PDM-deformed
harmonic oscillators of section 3, have their exact eigenvalues $E_{n}$ in
(66) and eigenfunctions $\Psi _{n}\left( q\left( x\right) \right) $ in (67).
This result (70) is, in fact, the most simplistic form of the PDM-Schr\"{o}%
dinger equation ever reported.

At this point, one should recollect that Cari\~{n}ena et al. \cite{34} have
started with the Killing vector fields for PDM geodesic motion (using a PDM
Lagrangian for quasi-free, $V\left( x\right) =0$) and have quantized the
associated Noether momentum (and not the canonical momentum) to come out
with a PDM Hamiltonian operator in exact accord with our $\hat{H}_{2}$ of
(68) to imply the PDM Schr\"{o}dinger equation (70). Yet, one should notice
that their PDM Noether momentum operator is reported as%
\begin{equation}
\hat{P}=-\frac{i}{\sqrt{2m\left( x\right) }}\partial _{x}=\frac{\hat{p}}{%
\sqrt{2m\left( x\right) }}
\end{equation}
where $\,\hat{p}=-i\partial _{x}$ is the regular textbook constant mass
momentum operator used in (69) and (70).

Finally, the PDM quantum supersymmetric approach with the asymptotic
geometrical classical oscillator correspondence by Cruz et al. \cite{21},
the factorization approach by Mustafa and Mazharimousavi \cite{27}, the
construction of the PDM-momentum operator approach by Mustafa and Algadhi 
\cite{10}, and the current PDM creation and annihilation oscillator
operators approach, all confirm and emphasize that the PDM-Hamiltonian $\hat{%
H}_{1}$ of (29) is the only surviving one out of the von Roos PDM
Hamiltonians. However, the quantization approach of the PDM Noether momentum
by Cari\~{n}ena et al. \cite{34}, and our analysis and discussions in the
current methodical proposal suggest that $\hat{H}_{2}$ of (68) is not only
correlated with $\hat{H}_{1}$ but also more simplistic user-friendly than $%
\hat{H}_{1}$ that of (63). It should replace the von Roos PDM-Hamiltonian
(1), therefore.\linebreak

\end{document}